\documentclass[pra,aps,twocolumn,superscriptaddress,showpacs]{revtex4-1}
\usepackage{braket}
\usepackage{soul}
\usepackage{hyperref}
\usepackage{dsfont}
\usepackage{amssymb}
\usepackage{amsmath}
\usepackage{amsfonts}
\usepackage{mathtools}
\usepackage{lipsum}
\usepackage{graphicx}
\usepackage{dcolumn}
\usepackage[dvipsnames]{xcolor}
\usepackage{bm}
\usepackage{multirow}
\graphicspath{{fig/}}
\newcommand{\affilITMO}{School of Physics and Engineering, ITMO University, St. Petersburg 197101,  Russia}
\newcommand{\school}{School №45, St. Petersburg,  Russia}
\newcommand{\schooll}{School №1589, Moscow,  Russia}
\usepackage{physics}
\usepackage{appendix}

\definecolor{linkcolor}{HTML}{0176ba}
\definecolor{urlcolor}{HTML}{0176ba} 
\definecolor{citecolor}{HTML}{900020}
\hypersetup{pdfstartview=FitH,  linkcolor=linkcolor,urlcolor=urlcolor, citecolor=citecolor, colorlinks=true}

\newcommand{\RMIN}{\widetilde{r}_{\mathrm{min}}} 

\begin{document}

\title[]{
Non-radiative configurations of a few quantum emitters ensembles: evolutionary optimization approach\\
}

\author{Ilya Volkov}
  \email{ilya.volkov@metalab.ifmo.ru}
 \affiliation{\affilITMO}

\author{Stanislav Mitsai}%

\affiliation{\affilITMO 
}%
\affiliation{\school 
}%

\author{Stepan Zhogolev}%

\affiliation{\affilITMO 
}%
\affiliation{\schooll
}%

\author{Danil Kornovan}

\affiliation{\affilITMO 
}%
\author{Alexandra Sheremet}

\affiliation{\affilITMO 
}%
\author{Roman Savelev}

\affiliation{\affilITMO 
}%

\author{Mihail Petrov}
\affiliation{\affilITMO
}%


\begin{abstract}
Suppressing the spontaneous emission in atomic ensembles is one of the topical problems in quantum optics and quantum technology. While many approaches are based on utilizing the subradiance effect in ordered quantum emitters arrays, the ensemble configurations providing the minimal spontaneous emission rate are yet unknown. In this work, we employ the differential evolution algorithm to identify the optimal configurations of a few  atomic ensembles that support quantum states with maximal radiative lifetime. We demonstrate that atoms tend to assemble mostly in quasi-regular structures with specific geometry, which strongly depends on the minimally allowed interatomic distance $r_{\mathrm{min}}$. While the discovered specific non-radiative realizations of small ensembles cannot be immediately predicted, there is particular correpondence to the non-radiative states in the atomic lattices. In particular, we have found that states inheriting their properties either from the bound states in the continuum or band-edge states of infinite lattices dominate across a wide range of $r_{\mathrm{min}}$ values. Additionally, we show that for small interatomic distances the linear arrays  with modulated spacing have the smallest radiative losses exponentially decreasing as the size of the ensemble increases.

\end{abstract}

\maketitle

The interest in protecting quantum states from the decoherence caused by spontaneous emission keeps driving theoretical~\cite{Sheremet2012, ScullyPRL2015, CaiPRA2016, Kornovan2016, jen_cooperative_2018, Plankensteiner2019, Zhang2020Feb, Fofanov2021} and experimental~\cite{pavolini_experimental_1985, devoe_observation_1996, guerin_subradiance_2016, Rui2020Jul, Ferioli2021} research in modern quantum optics since the pioneering work of R.~ Dicke~\cite{Dicke1954}. Among various platforms realizing quantum simulations and quantum computation, ensembles of cold atoms can be considered as one of the most prospective ones due to outstanding progress in their positioning in free space or in the proximity of nanophotonic structures~\cite{barredo_synthetic_2018,Corzo2016}. The control of spontaneous emission rate in atomic ensembles was demonstrated by arranging them in large regular one-dimensional arrays in free space~\cite{Asenjo-Garcia2017, Zhang2020Dec, Kornovan2021,Fayard2023} or near a waveguide~\cite{sutherland_collective_2016, Asenjo-Garcia2017Mar, Kornovan2019, Ke2019Dec, Zhang2020Feb, Poddubny2020}, as well as in two-dimensional arrays~\cite{facchinetti_storing_2016, bettles_cooperative_2015, Ballantine2020Apr}. Radiative decay rate in such structures can be significantly reduced due to formation of collective subradiant states, following either a polynomial~\cite{Zhang2020Feb, Kornovan2019, volkov_strongly_2023} or exponential~\cite{Asenjo-Garcia2017} dependence on the number of quantum emitters in the array $N$.
\begin{figure}[t]
    \centering
    \includegraphics[width=1\columnwidth]{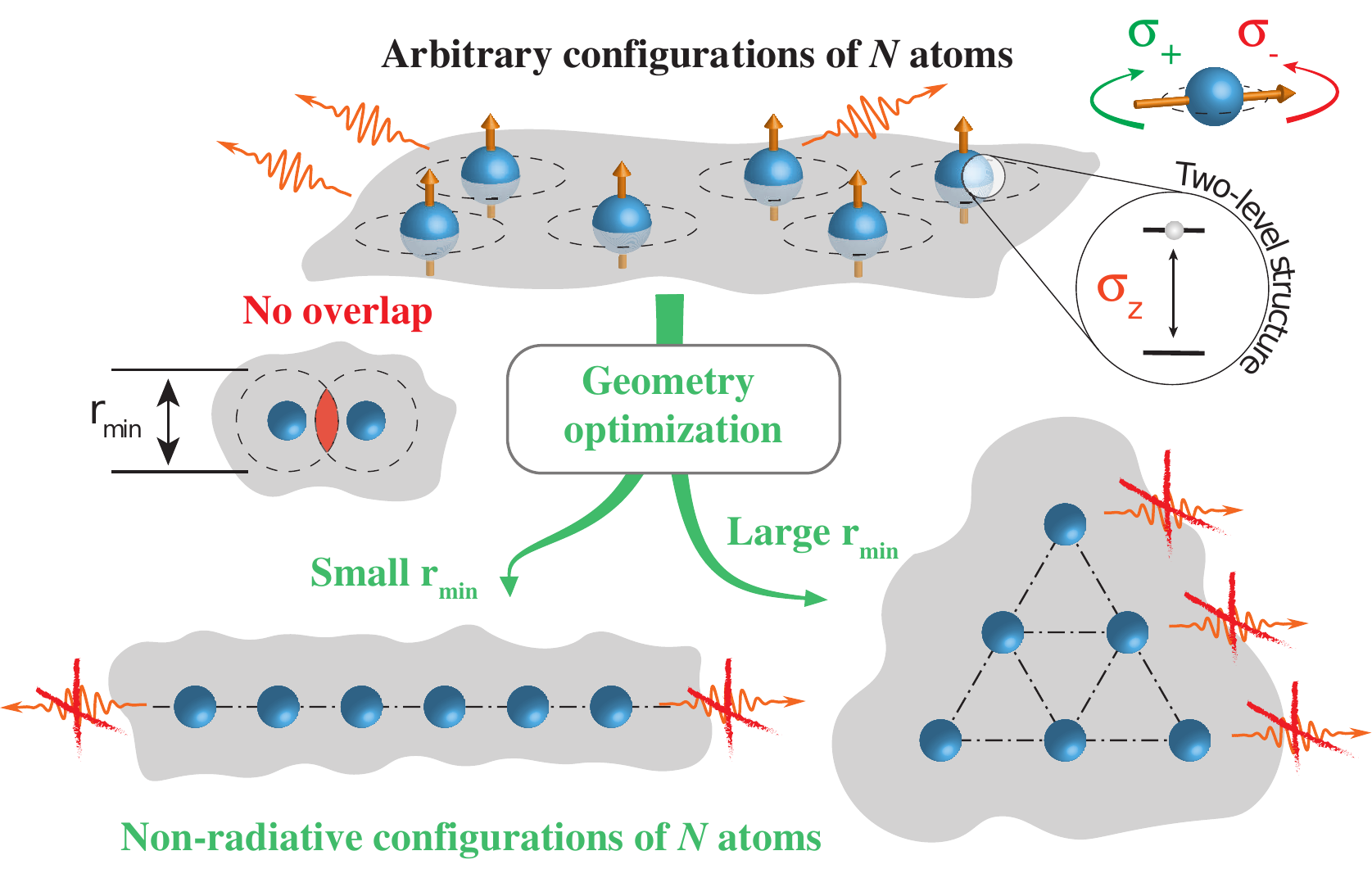}
    \caption{The quantum emitters having a two-level structure are arbitrary distributed in a plane $z=0$. The minimum allowed distance between the emitters is limited by $r_{\text{min}}$. Geometry optimization based on differential evolutionary method provides the configurations with minimal radiative losses. The final configuration strongly depends on the parameter $r_{\text{min}}$.}
    \label{fig:FIG1}
\end{figure}
Such findings, however, cannot be directly transferred onto the structures consisting of only a few atoms. The current approaches to engineering the radiative losses in small dipole ensembles are based on the symmetry of the arrays~\cite{de_paz_bound_2023, ustimenko_non-radiant_2023, volkov_strongly_2023}, with special attention to the rings composed of two-level atoms~\cite{freedhoff_cooperative_1986, cremer_polarization_2020, moreno-cardoner_subradiance-enhanced_2019, holzinger_nanoscale_2020, moreno-cardoner_efficient_2022,Holzinger2023Sep}, due to their high symmetry and relevance to various natural quantum systems. To this day, the general recipe for tackling the problem of loss suppression in ensembles of dipole emitters hos not been proposed. One of the main reasons is the large dimensionality of the parameter space, which leads to the high complexity of the problem rapidly growing with increasing the number of emitters. In this context, modern powerful numerical optimization algorithms could serve as reliable tools to efficiently solve the problem directly and uncover the importance of various physical mechanisms responsible for the formation of subradiant states.

Numerical algorithms based on evolutionary and machine learning approaches have already proved their efficiency in optimization of nanophotonic structures such as  subwavelength resonators  and metasurfaces ~\cite{krasikovOEA2022,pan_inverse_2022,mikhailovskaya_superradiant_2022,gladyshev_inverse_2023, ustimenko_multipole_2021} and for engineering quantum systems \cite{tranter_multiparameter_2018, rozenberg_inverse_2022}. In this Letter, we utilize the differential evolution method to numerically find the optimal planar atomic configurations that support quantum excitations with minimal spontaneous emission rate. The results of the simulations reveal that the atoms mostly tend to arrange in regular or quasi-regular structures. However, in some cases, the specific geometry turns out to be counterintuitive and difficult to predict based on the general physical principles established for large atomic arrays. 

\begin{figure}[t]
    \centering
    \includegraphics[width=1\columnwidth]{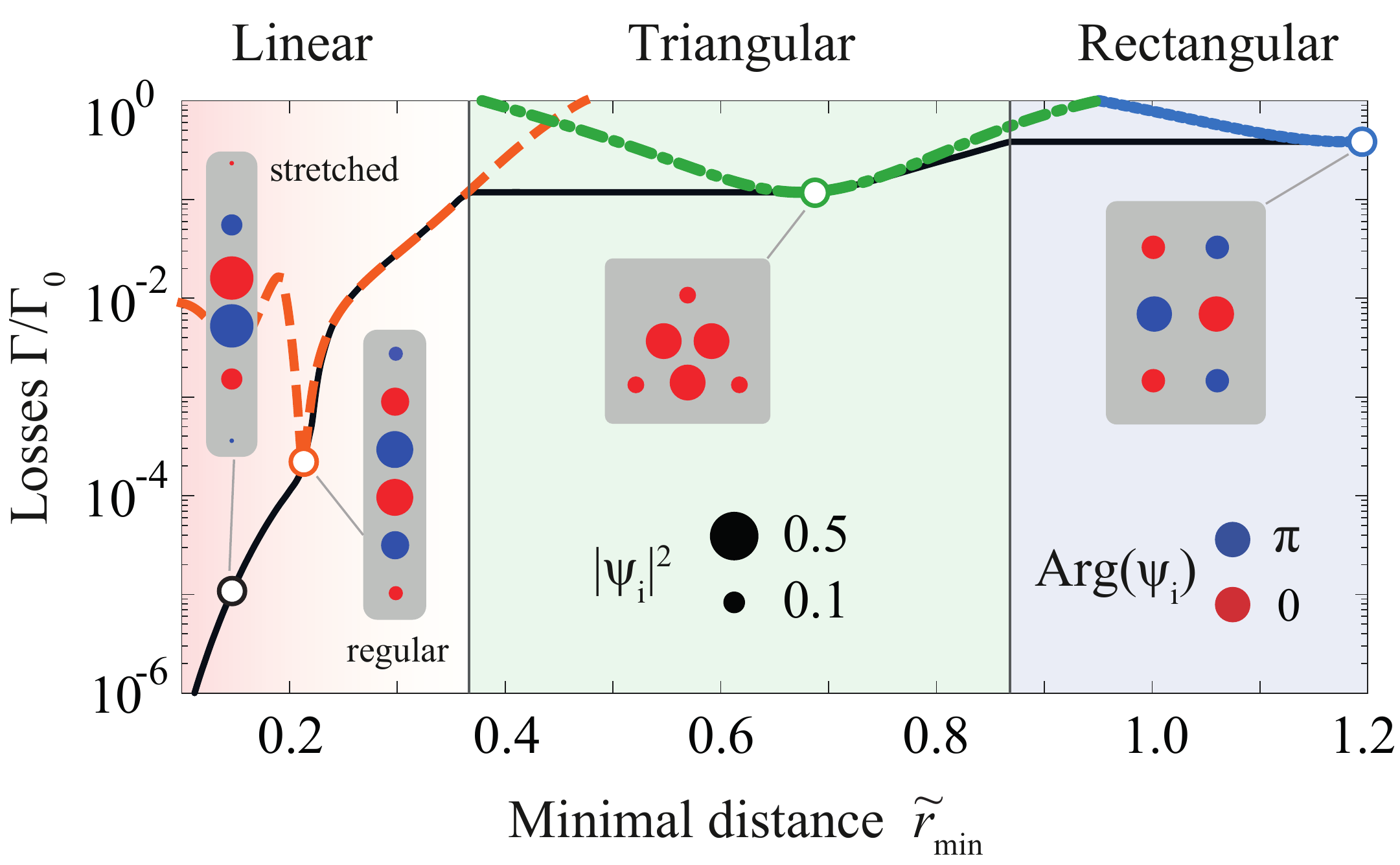}
    \caption{Calculated optimal atomic configurations and corresponding radiative losses for different minimum interatomic distance $\RMIN$ (black solid curve). Dashed orange, dashed-dotted green and dotted blue curves correspond to the radiative losses of the regular chain, triangle and rectangle, respectively, with a varied lattice constant. Wavefunctions $\psi$ of eigenstates corresponding to different distinct regions of $\RMIN$ are illustrated in insets with size of the dots being proportional to $|\psi|^2$ and color indicating the phase of the wavefunction $\rm{Arg}(\psi_i)$.}
    \label{fig:FIG2}
\end{figure}

The problem of finding the configuration of atoms with minimal radiative losses has, of course, a global yet unrealistic solution when all atoms tend to be located at one point in space. In a single excitation domain, this leads to the formation of subradiant states with radiative losses approaching zero~\cite{Dicke1954}. At the same time, the modern methods of quantum emitters' positioning have certain limitations on the minimal interatomic distance \cite{srakaew_subwavelength_2023, rui_subradiant_2020}. To account for that, the distance between the atoms $r$ was subject to constraint $r>r_{\text{min}}$ (see Fig.\ref{fig:FIG1}) during the optimization procedure. We also assumed throughout the paper that all atoms are located in the $z=0$ plane, thus limiting all possible configurations to two-dimensional ones.



In our model, all atoms are assumed to have {a two-level structure shown in Fig.~\ref{fig:FIG1}(a)}. In what follows, we mostly focus on the $\sigma_z$ dipole transition, assuming that it is independent from the other ones. This can be achieved, for instance, by applying a strong external magnetic field that spectraly isolates transitions. Case of atoms with $\sigma_\pm$ isolated transition is proceeded in  Supplementary Materials. The transition resonant frequency is denoted as $\omega_0$ and the rate of spontaneous emission as $\Gamma_0 = k_0^3 |\mathbf{d}|^2 / \left( 3 \pi \hbar \varepsilon_0\right)$, where $k_0 = \omega_0/c = 2 \pi/\lambda_0$ is the wavenumber in vacuum, and $\varepsilon_0$ is the vacuum permittivity. The coupling between the atoms is governed by the dipole-dipole interaction, described by the electromagnetic free space Green's tensor $\mathbf{G}(\mathbf{R},\omega)$ \cite{Novotny2012Sep}. The quantum states of the atomic ensemble are then described by the effective Hamiltonian~\cite{Markel2007Feb, AsenjoPRA2017}:

\begin{multline}     
\hat{\mathcal{H}}_{\text{int}}  = \hbar \Bigl( \omega_0 - \frac{i}{2}\Gamma_0 \Bigr) \sum\limits_{i=1}^N \hat{\sigma}_i^{\dag} \hat{\sigma}_i -\\- \sum\limits_{i\neq j} \frac{3\pi \hbar\Gamma_0}{k_0} \mathbf{e_d^*} \cdot \mathbf{G}(\mathbf{R}_{ij})\cdot \mathbf{e_d} \ \hat{\sigma}_i^{\dag} \hat{\sigma}_j,
\label{Hamiltonian}
\end{multline} 
where $\hat{\sigma}_i^{\dag}$ ($\hat{\sigma}_i$) is the creation (annihilation) operator for the excitation on the $i$-th emitter, $\mathbf{e_d} = \mathbf{d}/d$ is a unit vector of atomic transition dipole moment. The first term in Eq.~\eqref{Hamiltonian} is responsible for the spontaneous emission of single atom in vacuum, while the second one describes the dipole-dipole interactions between the atoms. Here, we apply the Markovian approximation $\mathbf{G}(\mathbf{R},\omega) \approx \mathbf{G}(\mathbf{R},\omega_0)$, which is justified by a very narrow resonance of a single atom $\Gamma_0 \ll \omega_0$. The radiative decay rate of the $j$-th collective atomic excitation is determined by the imaginary part of the complex eigenfrequency $\omega_j - i \Gamma_j/2$ of the non-Hermitian Hamiltonian~\eqref{Hamiltonian}, where $j=1 \ldots N$. The minimal decay rate $\text{min}(\Gamma_j)$ then serves as a goal function for the optimization, with the coordinates of the atoms being the input arguments.

A differential evolution method~\cite{Storn1997Dec} employed in our work applies ideas of genetic algorithm to a set of continuous parameters. In the case of an ensemble of $N$ atoms located in a plane, there are $2N-3$ degrees of freedom (two coordinates of $N-1$ atoms minus one due to invariance of the eigenstates to the rotation of the whole structure). This significantly limits the maximal size of the optimized structures, which in our work was equal to $N=8$. We have maintained two general restrictions during the optimization. The first one is the lower limit of the interatomic distance $r_{\mathrm{min}}$, see Fig.~\ref{fig:FIG1}. The second one is the upper limit on the overall size of the structure, ensuring that it is confined inside a circle with a radius of $5 \lambda_0$.

The results of the optimization procedure for $N=6$ $z$-polarized dipole transitions are depicted in Fig.~\ref{fig:FIG2} showing the dependence of the minimal normalized radiative losses $\Gamma/\Gamma_0$ on the minimal interatomic distance, normalized by the dipole transition wavelength $\RMIN = r_{\mathrm{min}}/\lambda_0$ (black solid line). One may also find similar results for circular polarized dipoles in Supplementary Materials. For the considered range $\RMIN \in [0.1,1.2]$, the optimal configurations can be separated into four types shown in the insets; the circles correspond to the positions of the atoms, area of the circles is proportional to the square amplitudes of the excitation at each atomic site, while the color indicates the phase of this excitation. It turns out that for small minimal distances $\RMIN \lesssim 0.35$ linear configurations are the optimal ones. In the region $0.22 \lesssim \RMIN \lesssim 0.35$, the quantum emitters are (almost) regularly distributed in space. To be more precise, all distances between the neighboring dipoles differ by less than $\sim 10^{-4} \lambda_0$. For the separation distance $\RMIN \lesssim 0.22$, the optimal configuration becomes stretched, with the distance between the emitters increasing from the center to the end of the chain.

\begin{figure}[t]
    \centering
    \includegraphics[width=1\columnwidth]{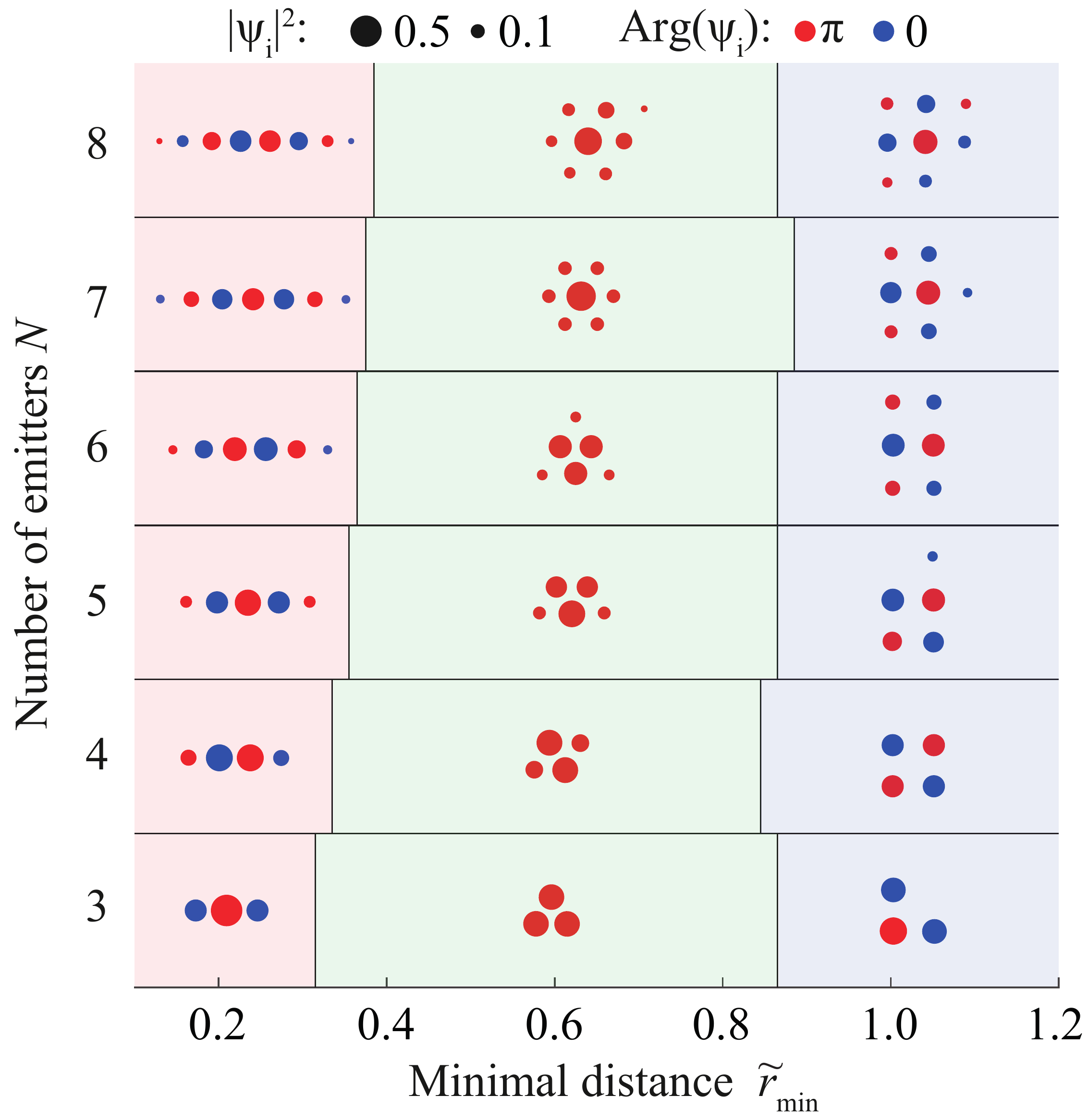}
    \caption{Wavefunction of the most subradiant eigenstates in the optimal configurations for various restricting distances $\RMIN$ and number of emitters $N$. Size of the dots is proportional to $|\psi|^2$ and color indicates the phase of the $\psi$.}
    \label{fig:FIG3}
\end{figure}

For larger $\RMIN$ in the range $ 0.38 \ldots 0.88$, a quasi-regular triangular shape is formed. One should note that due to a large distance between the emitters of around $\approx \lambda_0/2$ and corresponding $\approx\pi$ phase shift between the wavefunction in the neighbor emitters, the minimal losses are observed when excitation at all emitters has the same phase. For even larger distances $\RMIN \approx 0.88 \ldots 1.2$, a rectangular array arranged in a square lattice with the out-of-phase eigenstate wavefunction exhibits minimal losses. Note that in the most range of $\RMIN$ values the optimal configurations are close to the regular ones. Dashed orange, dashed-dotted green and dotted blue curves in Fig.~\ref{fig:FIG2} show the radiative losses of the most subradiant states supported by three characteristic realizations of regular structures: chain, triangle, and rectangle, respectively. The $\RMIN$ for these curves denotes not just the minimal distance between the atoms but rather the lattice constant. One can observe that the difference between the regular and optimal configurations is rather small for the most part of the $\RMIN$ values.

The typical geometrical configurations that minimize the radiative losses remain similar for different number of emitters $N=3 \ldots 8$, as shown in Fig.~\ref{fig:FIG3}. For small minimal distances, the optimal subradiant configurations are linear arrays with collective eigenstates characterized by out-of-phase amplitudes of excitation in the neighboring atoms and maximal amplitude at the center of the chain. These states inherit their subradiant properties from the staggered guided modes at the band edge of the infinite periodic chains~\cite{WeberPRB2004,FigotinPRE2005}. The appropriate modulation of the distances between the emitters leads to the stronger localization of the wavefunction in the center of the chain (see e.g. left inset in Fig.~\ref{fig:FIG2}) and consequently to much weaker radiative losses. Such an effect tends to be more pronounced for the ensemble of larger sizes.

For larger separations, $\RMIN \approx 0.4 \ldots 0.8$ atoms tend to arrange in a quasi-regular triangular lattice. The in-phase amplitude of the excitation on all atoms links such states to the so-called bound state in continuum (BIC) -- non-radiative states that appear for an infinite lattice in the center of the Brillouin zone~\cite{Hsu2016Jul,Sadrieva2019Sep,OvervigPRB2020}. Interestingly, while the BIC states exist in lattices of a wide class of symmetries~\cite{OvervigPRB2020, Sadrieva2019Sep}, the most subradiant states in small finite structures assemble in a triangular lattice. For even larger separation distances $\gtrsim 0.8$ the atoms weakly interact with each other. Consequently, the strongest suppression of losses is achieved for the out-of-phase amplitudes of the wavefunction in the neighbor atoms in a fragment of the square lattice. Similar results for circularly polarized emitters are presented in Supplementary Materials.

\begin{figure*}[t!]
    \centering
    \includegraphics[width=2\columnwidth]{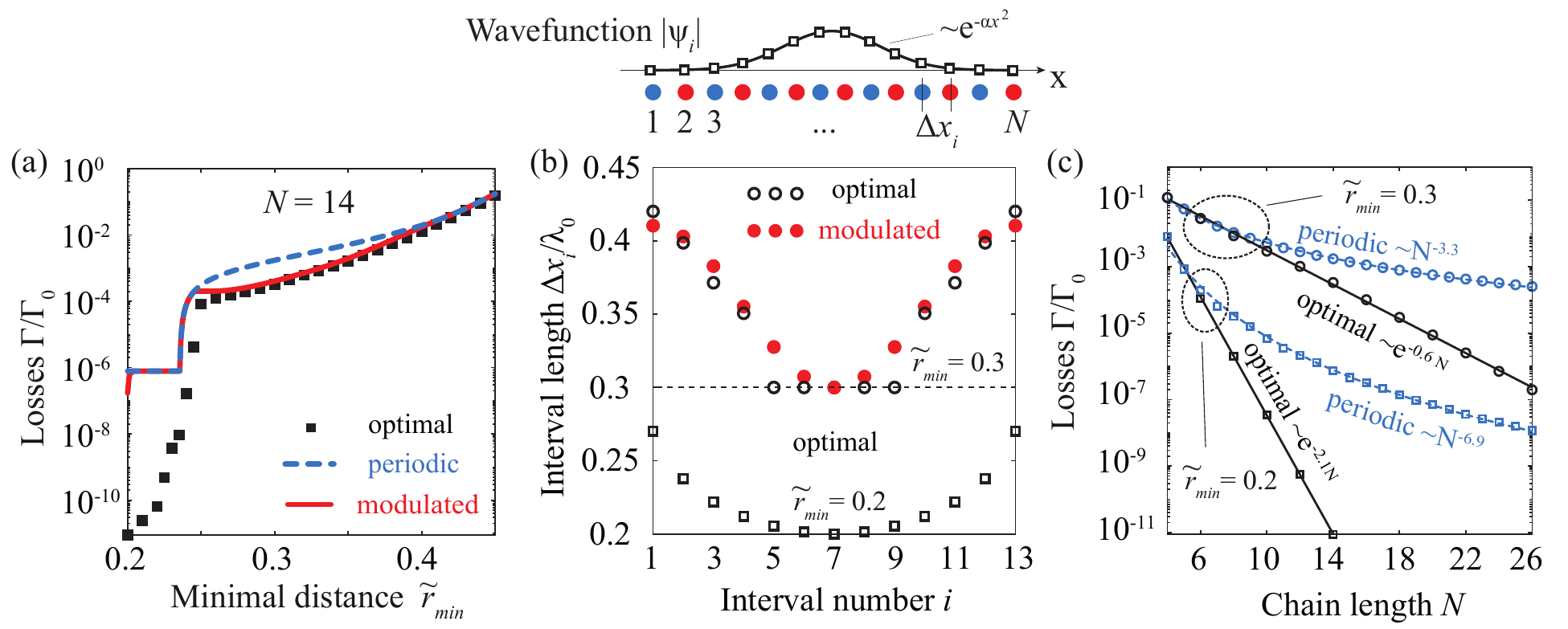}
    \caption{Inset (top): distribution of the excitation across the quantum emitters. (a) Dependence of losses on restricted distance $\RMIN$ for $N=14$ atoms in cases of optimized free configurations (black squares),  periodic chain (blue dashed), a chain with sinusoidal modulation (red solid line). (b) Distances between neighboring atoms in cases of optimized configurations and linear array with sinusoidal modulation for $N=14$ atoms with $\RMIN =0.2 (0.3)$.(c) Scaling of losses for optimal and periodic geometries as function of the array size for $\RMIN =0.2 (0.3)$. }
    \label{fig:FIG4}
\end{figure*}
    
The linear arrays with minimal losses for short interatomic spacing deserve special attention due to strong interest in waveguide coupled quantum ensembles  \cite{Sheremet2023Mar} and many already published reports devoted to such geometry~\cite{Asenjo-Garcia2017, Kornovan2016, Kornovan2019, Kornovan2021, pan_inverse_2022,  ZhangPRL2019, Zhang2020Dec, PichuginNanophotonics2021}. The calculations presented in Figs.~\ref{fig:FIG2},\ref{fig:FIG3} have revealed that for $N\leq 8$ the optimal 1D configurations gradually change from the regular chains to the stretched ones with the decrease of the $\tilde r_{\mathrm{min}}$. By restricting the geometry to only 1D configurations, we were able to extend these results up to $N=26$ atoms. We have also increased the maximal radius of the circle that confines the structure to $N(r_{\mathrm{min}}+\lambda_0)/2$. In order to analyze the exact profiles of atom positions along the chain, in Fig.~\ref{fig:FIG4} (a) we plot radiative losses of the states in optimized chains (black squares) along with the regular chains (dashed blue curve) and the chain with modulated interatomic distances (solid red curve) for a fixed number of atoms $N=14$. Modulation was chosen according to the following phenomenological formula $\Delta x (i) = r_{max}-(r_{max}-r_{\mathrm{min}}) \text{sin}^2({\pi (i-1)}/({N-2}))$, where $i = 1..N-1$ is the number of the interval between atoms. The parameters of the regular and modulated chains were also optimized, subject to a constraint on the minimal interatomic spacing.

One can observe that the modulated chain  matches well with  the optimal configuration in the range $\RMIN \gtrsim 0.25$. This is illustrated in Fig.~\ref{fig:FIG4}(b) with black and red circles showing the spatial profile of the interatomic spacing for $\RMIN=0.3$ for the optimal configuration and for the modulated one, respectively. As seen from Fig.~\ref{fig:FIG4} (a), the difference with the regular chain in this range of $r_{\mathrm{min}}$ reaches an order of magnitude for a given size of the chain $N=14$. For a larger number of atoms, however, the difference becomes much more pronounced due to qualitatively different dependence of the radiative losses on the size of the structure, as shown in Fig.~\ref{fig:FIG4}(c), while the regular chain exhibits well-known $\sim N^{-3}$ dependence, the losses in a modulated one decrease exponentially. For smaller $\RMIN \lesssim 0.25$, the losses decrease much faster both with the decrease of the $\tilde r_{\mathrm{min}}$ or with the increase of $N$, see Fig.~\ref{fig:FIG4}(a,c). Such a change is accompanied by a qualitative change in the interatomic spacing modulation. As shown in Fig.~\ref{fig:FIG4}(b) with black squares, the modulation profile becomes close to the parabolic one. {We note while exponential scaling of the radiative losses in modulated 1D arrays was already demonstrated  in \cite{Asenjo-Garcia2017}, here  we advance these  results by providing the most optimal configurations.}

Importantly, the results obtained in our work for an ensemble of a few dipole emitters cannot be easily predicted based on the known physical mechanisms of radiative loss suppression established for larger arrays. It is known that the radiative losses of the band-edge states in regular linear arrays can decrease with the atom number as fast as $N^{-7}$~\cite{Kornovan2019,Kornovan2021}, and in regular two-dimensional arrays as fast as $N^{-5}$~\cite{volkov_strongly_2023}. At the same time, quasi-BIC states exhibit much slower dependence of losses on the size $N^{-1}$~\cite{Facchinetti2016Dec}. Nevertheless, in small ensembles, quasi-BIC states appear to be more subradiant in a wide range of $\RMIN$ values, while band-edge states in 1D chains dominate only for small enough interatomic distances. For large distances $\RMIN \rightarrow 1$ the opening of the diffraction channels makes the competition between the aforementioned mechanisms of the subradiance even more complicated and the results of the optimization even less predictable.



In conclusion, we utilized the differential evolution algorithm to find the optimal configurations of small atomic ensembles that support eigenstates with minimal radiative losses - subradiant states. Depending on the minimum allowed distance between the atoms, different quasi-regular structures are inclined to form, such as one-dimensional chains, fragments of triangular, and square lattices. Due to the various geometries and loss suppression mechanisms in these structures, it is not trivial to predict in advance the exact optimal configurations.
\\
\section*{Acknowledgement}
The authors acknowledge Kseniia Baryshnikova for her efforts in making this work happen, and Nikita Ustimenko for fruitful discussions. The work was supported by the Federal Academic Leadership Program Priority 2030.

\bibliographystyle{ieeetr}
\bibliography{ref}

\end{document}


\title{Non-radiative configurations of a few quantum emitters ensembles: evolutionary optimization approach:\\Supplementary Material}

\author{Ilya Volkov}
  \email{ilya.volkov@metalab.ifmo.ru}
 \affiliation{\affilITMO}

\author{Stanislav Mitsai}%

\affiliation{\affilITMO 
}%
\affiliation{\school 
}%

\author{Stepan Zhogolev}%

\affiliation{\affilITMO 
}%
\affiliation{\schooll
}%

\author{Danil Kornovan}

\affiliation{\affilITMO 
}%
\author{Roman Savelev}

\affiliation{\affilITMO 
}%

\author{Mihail Petrov}
\affiliation{\affilITMO
}%



\maketitle
\tableofcontents
\clearpage

\section*{A. Implementation of differential evolution method }


We start with an arbitrary 2D system of $N$ atoms. Without loss of generality the first atom is placed at the origin,  $(0,0)$ point, in polar coordinates $(\rho,\phi)$, and the second one is placed at the point $(r_2,0)$. Consequently, there are $2N-3$ arbitrary atomic coordinates which we consider as components of a parameter vector $\overline{\bf{v}}$ for the optimization problem. The boundaries of the parametric space $\overline{\bf{v}}$ are defined indirectly through the lower limit of interatomic distances $r_{min}$ and the upper limit on the distance from each of the atoms to the origin $N (r_{min}+\lambda_0)/2 $. In our case, the objective function for global minimization is the decay rate minimal, min($\Gamma_j$),  which we find by the direct diagonalization of the Hamiltonian from Eq.~(1), among all $N$ collective modes for a fixed atomic configuration.

As soon as the objective function is obtained fully numerically, the optimization problem in general is not differentiable, as is required by classical optimization methods such as gradient descent. Moreover, since it is not possible to extract $2N-3$ parameters from the Hamiltonian before diagonalization, this optimization problem belongs to the NP class of problems in terms of complexity. Such problems are not solvable in polynomial time and are therefore usually treated as black box problems. Therefore, we applied a differential evolution (DE) approach \cite{Storn1997Dec}, based on stochastic and iterative improvement of candidate solutions, to minimize the objective function. The algorithm starts with $p$ random initial vectors in parametric space, i.e. initial population  ${\bf{V}}_{0}$, where $i=1..p$. The number of vectors $p$ in each population is defined manually, in current research it was at least twice the length of the parameter vector $\overline{\bf{v}}$ size, $2(2N-3)$, but not less than $10$.

Next, the DE algorithm itself begins with iterative mutation, crossover, and selection applied to initial and consecutive generations of parameter vectors. Iteration for candidate vector $\overline{\bf{v}}_{i,g}$ in generation $g = 1..G$ begins with choosing the best parameter vector $\overline{\bf{v}}_{g}$ in population ${\bf{V}}_{g}$. Starting with a mutation, a difference of two random vectors in the current population, $\overline{\bf{v}}_{j,g}$ and $\overline{\bf{v}}_{k,g}$ is added to $\overline{\bf{v}}_{g}$ with some weight:
\begin{equation}
\overline{\bf{v}}_{i,g+1}' = \overline{\bf{v}}_{g} + F_g(\overline{\bf{v}}_{j,g} - \overline{\bf{v}}_{k,g}),
\label{Mutation}
\end{equation} 
where $F_g$ is a mutation constant that randomly changes from one generation to another in the interval $[0.5,1]$. Next, a crossover between the mutant vector $\overline{\bf{v}}_{i,g+1}'$ and the original candidate $\overline{\bf{v}}_{i,g}$ begins by sequentially and probabilistic filling of the trial vector $\overline{\bf{v}}_{i,g+1}''$ by either components of $\overline{\bf{v}}_{k,g+1}'$ or $\overline{\bf{v}}_{g}$. At the end, selection is applied: the trial vector is compared with the original candidate in terms of the objective function value. The best vector occupies $\overline{\bf{v}}_{i,g+1}$ in the next generation. And finally, the trial vector is analogously played off with the best original candidate $\overline{\bf{v}}_{g}$ in generation.

The iteration described above is applied to all vectors in the population $\overline{\bf{v}}_{g}$ for all generations $g=1..G$ and stops when the standard deviation of vectors in the current generation is small enough. In our case, the algorithm stops at generation $G$ when $\sigma({\bf{V}}_{G}) \lesssim 0.01 \langle{\tilde{\Gamma}}_G\rangle$, where $\langle{\tilde{\Gamma}}_G\rangle$ is the average value by population of objective function values normalized to $1$.

\section*{B. Optimal configurations for the case of $\sigma_{\pm}$ polarization.}

Similarly to $\sigma_z$ dipole transitions, one can consider two-level atoms with $\sigma_\pm$ dipole transitions. For this purpose, instead of ${\bf{e_d}} = {\bf{e}}_z$ we substitute ${\bf{e_d}} = ({\bf{e}}_x \pm {i\bf{e}}_y)/\sqrt{2}$ in Eq.~[1] from the main text, where ${\bf{e}}_{x,y,z}$ are the unit basis vectors. Thereby, here we provide results for circular polarized dipoles in a manner similar to results for linear dipoles in the main text.

\begin{figure}[h!]
    \centering
    \includegraphics[width=1\columnwidth]{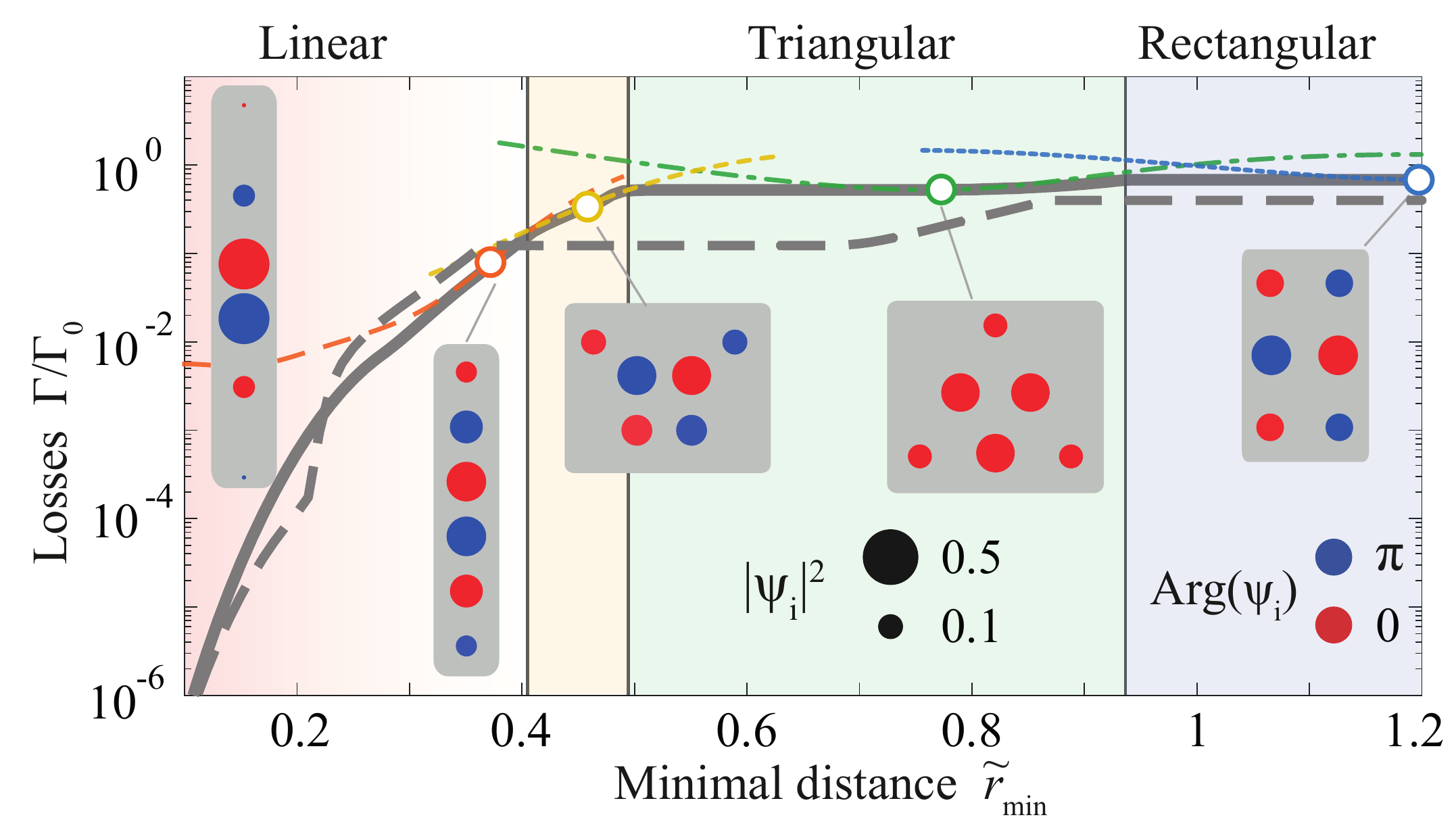}
    \caption{Calculated optimal atomic configurations and corresponding radiative losses for different minimum interatomic distance $\RMIN$ for $\sigma_\pm$ and $\sigma_z$ transitions (grey solid and grey dashed curve). Dashed orange, dashed yellow, dashed-dotted green and dotted blue curves correspond to the radiative losses of specified structures with a varied lattice constant. Wavefunctions $\psi$ of eigenstates corresponding to different distinct regions of $\RMIN$ are illustrated in insets with size of the dots being proportional to $|\psi|^2$ and color indicating the phase of the wavefunction $\rm{Arg}(\psi_i)$.}
    \label{fig:S1}
\end{figure}

Firstly, for $N=6$ dipoles we established optimal configurations and their radiative losses dependencies on $\RMIN$. As one may see in Fig.~\ref{fig:S1}, for $\sigma_\pm$ dipoles we obtained all optimal geometries we have already seen for $\sigma_z$ dipoles. Additional to that, a new optimal configuration was obtained for $0.41\leq \RMIN \leq 0.49$, where out-of-phase dipoles form a non-trivial 2D structure. One may also notice that for most values of $\RMIN$, configurations of $\sigma_\pm$ dipoles are more radiative than $\sigma_z$ dipoles configurations.

\begin{figure}[t!]
    \centering
    \includegraphics[width=1\columnwidth]{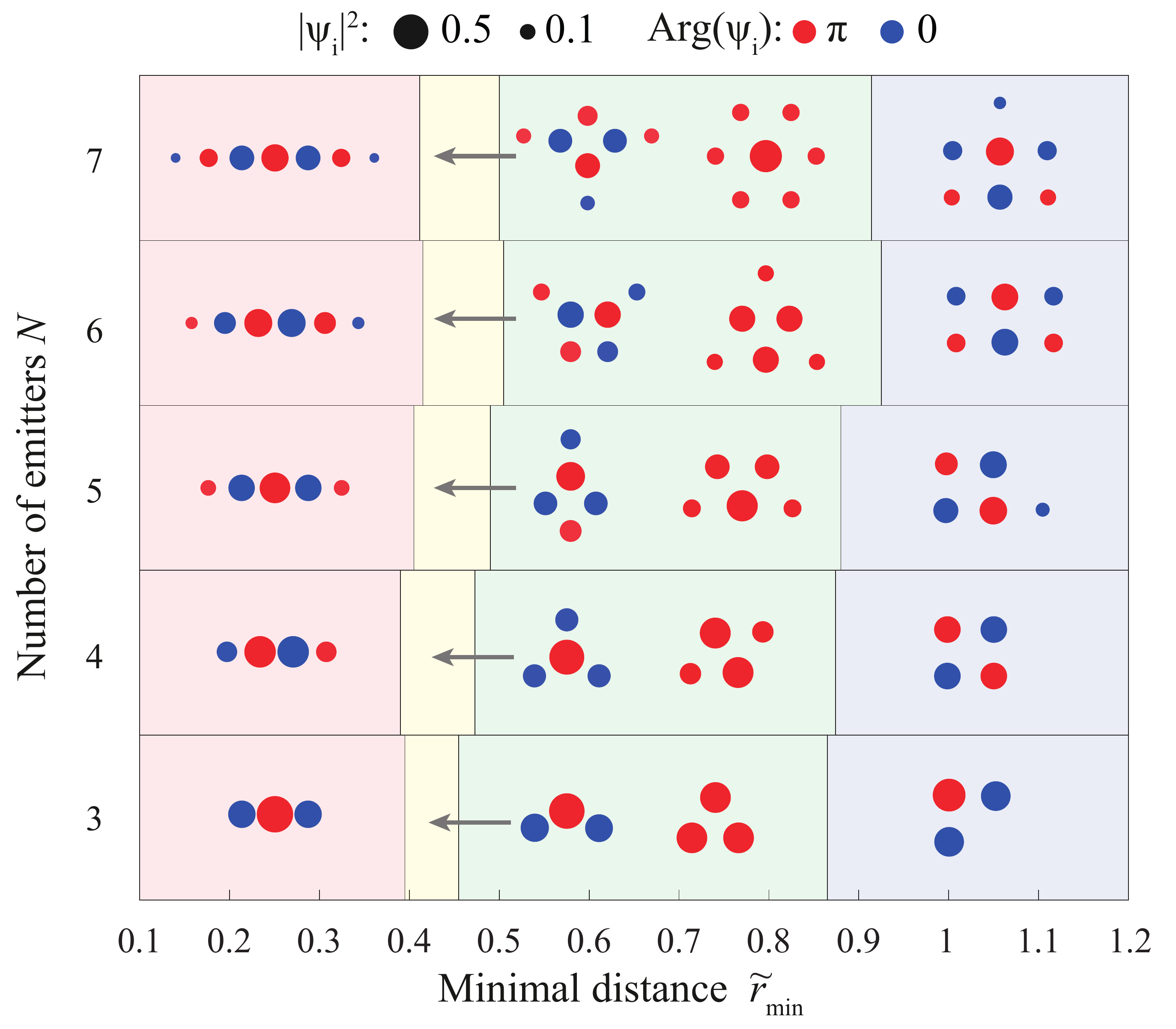}
    \caption{Wavefunctions of the most subradiant eigenstates in the optimal configurations for various restricting distances $\RMIN$ and number of emitters $N$. Size of the dots is proportional to $|\psi|^2$ and color indicates the phase $\rm{Arg}(\psi_i)$.}
    \label{fig:S2}
\end{figure}

Secondly, we calculated optimal configurations for $N=3..7$ dipoles, see Fig.~\ref{fig:S2}. All optimal geometries obtained for the case of $\sigma_z$ dipoles were revealed again for $\sigma_\pm$ dipoles. In addition, for $\RMIN \sim 0.4$ new 2D geometries composed of out-of-phase dipoles were obtained.

\bibliographystyle{ieeetr}
\bibliography{ref}